\documentclass[]{llncs}

\usepackage{algorithm}
\usepackage{algorithmicx}
\usepackage[noend]{algpseudocode}

\usepackage{graphicx}
\usepackage{amsmath}
\usepackage{amssymb}

\usepackage{subfig}

\usepackage{hyperref}

\title{Scalable Multiagent Coordination with Distributed Online Open Loop Planning}
\author{Lenz Belzner \and Thomas Gabor}
\institute{Institute for Informatics\\Mobile and Distributed Systems\\LMU Munich}

\begin{document}

\maketitle

\begin{abstract}
We propose distributed online open loop planning (DOOLP), a general framework for online multiagent coordination and decision making under uncertainty. DOOLP is based on online heuristic search in the space defined by a generative model of the domain dynamics, which is exploited by agents to simulate and evaluate the consequences of their potential choices.
We also propose \textit{distributed online Thompson sampling} (DOTS) as an effective instantiation of the DOOLP framework. DOTS models sequences of agent choices by concatenating a number of multiarmed bandits for each agent and uses Thompson sampling for dealing with action value uncertainty. The Bayesian approach underlying Thompson sampling allows to effectively model and estimate uncertainty about (a) own action values and (b) other agents' behavior. This approach yields a principled and statistically sound solution to the exploration-exploitation dilemma when exploring large search spaces with limited resources.
We implemented DOTS in a smart factory case study with positive empirical results. We observed effective, robust and scalable  planning and coordination capabilities even when only searching a fraction of the potential search space.
\end{abstract}

\section{Introduction}

We present a framework for efficient and scalable online multiagent coordination and decision making under uncertainty. We present distributed online open loop planning (DOOLP), a general framework based on online heuristic search in the space defined by a generative model of the domain dynamics, which is exploited by agents to simulate and evaluate the consequences of their potential choices.

We also present a particular instantiation of DOOLP, \textit{distributed online Thompson sampling} (DOTS). DOTS models sequences of agent choices by concatenating a number of multiarmed bandits for each agent. In the distributed sequential setting, optimal bandit choices depend on subsequent action value estimates and preferences of other agents. We propose to use Thompson sampling as an efficient technique to estimate the value of actions by simulation. We show that the Bayesian approach underlying Thompson sampling allows to effectively model and estimate the uncertainty about (a) own action values and (b) other agents' behavior. DOTS uses probabilistic sampling strategies that are updated in a Bayesian way based on simulated reward in order to solve the exploration-exploitation dilemma when exploring large search spaces with limited resources.

We implemented DOTS in a smart factory case study with more than $10^{22}$ states, and showed its applicability to distributed constraint optimization problems. We observed effective problem solving and coordination capabilities even when only searching a fraction of the potential search space. We compared different action selection strategies for dealing with the exploration-exploitation dilemma, and observed that Bayesian selection performed more effectively than other baseline approaches. We also evaluated the effect of gossip-like coordination when planning and observed that communication does not have to be global in order to maintain stable and effective global coordination results.

The remainder of the paper is structured as follows. Section \ref{sec:related} outlines related work. Section \ref{sec:doolp} describes the DOOLP framework for distributed online open loop planning. In Section \ref{sec:dots} we discuss the DOTS algorithm as a Bayesian instantiation of DOOLP. We present an empirical case study in Section \ref{sec:results} and discuss our results. We conclude and outline venues for further research in Section \ref{sec:conclusion}.

\section{Related Work}
\label{sec:related}

In this section, we recap decentralized Markov decision processes, online planning, open loop planning, multiarmed bandits and Thompson sampling.

\subsection{Decentralized MDPs}

A finite decentralized Markov decision process $\mathcal{M}$ (DecMDP) is defined by a tuple
$\mathcal{M} = (\textit{Ag}, S, s_0, \{A_\textit{ag}\}, P(S | S, A), R, h)$, where
\begin{itemize}
	\item $\textit{Ag}$ is a set of agents,
	\item $S$ is a set of states,
	\item $s_0 \in S$ is an initial state,
	\item $A_\textit{ag}$ is the set of actions of an agent $\textit{ag} \in \textit{Ag}$,
	\item $A = \otimes_{\textit{ag} \in \textit{Ag}} A_\textit{ag}$ is the set of joint actions,
	\item $P(S | S, A)$ is the transition distribution, a probability distribution over (successor) states when executing a joint action in a given (preceding) state,
	\item $R : S \times A \times S \rightarrow \mathbb{R}$ is a reward function,
	\item $h \in \mathbb{N}^+$ is a planning horizon.
\end{itemize}

At every step $t < h$ in state $s_t \in S$, each agent $\textit{ag} \in \textit{Ag}$ executes an action $a_\textit{ag} \in A_\textit{ag}$, yielding the joint action $a_t = \bigcup_{\textit{ag} \in \textit{Ag}} a_\textit{ag}$. The DecMDP then preforms a transition w.r.t. the transition distribution, and an immediate reward $R(s_t, a_t, s_{t + 1})$ is observed.

Solving a DecMDP consists in finding a policy $\Pi_\textit{ag} : P(A_\textit{ag} | S)$ for each agent $\textit{ag} \in \textit{Ag}$ such that some optimization criterion is satisfied when executing the set of policies. Note that joining the individual agents' policies yields a joint policy $\Pi : P(A | S)$.

A possible optimization criterion for finite DecMDPs is the expected cumulative reward. The cumulative reward $\mathrm{CR}$ of a DecMDP is defined as the sum of rewards observed when executing the policies of a set of agents.
$$\mathrm{CR}(s_0, a_0, s_1, a_1, ..., s_{h - 1}, a_{h - 1}, s_h)  = \sum_{t = 0}^{h - 1} R(s_t, a_t, s_{t + 1})$$

Optimizing the expected cumulative reward is then equal to maximizing the expectation of the cumulative reward.
$$\max \mathbb{E}\left[ \sum_{t = 0}^{h - 1} R(s_t, a_t, s_{t + 1}) \right]$$

We refer to \cite{oliehoek2016concise} for an in-depth discussion of decentralized Markov decision processes.

\subsection{Online Planning}

\begin{figure}
	\centering
	\includegraphics[width=0.5\textwidth]{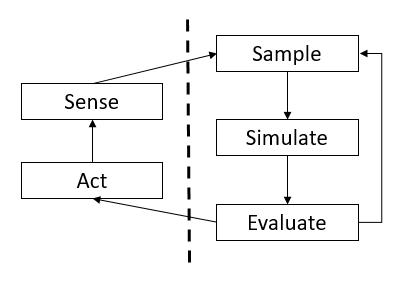}
	\caption{Schematic representation of online planning. Note the two feedback loops of execution (left) and planning (right). The dashed vertical line indicates physical (left) and cognitive (right) components of an online planning agent.}
	\label{fig:olp}
\end{figure}

Online planning, or local search, repeatedly interleaves a planning loop with an execution loop \cite{weinstein2013open,belzner2015onplan}. An online planning agent requires a probabilistic generative model of its domain dynamics, such as a transition distribution of a DecMDP. Online planning consists of two loops with different frequencies.
\begin{itemize}
	\item A high-frequency planning loop samples a plan from a stochastic policy, simulates its consequences w.r.t. some optimization objective (e.g. the expected cumulative reward) and updates the policy in order to increase the probability of generating useful plans w.r.t. the given notion of utility.
	\item A low-frequency execution loop consists of sensing the current state of the environment, planning by repeating the inner loop multiple times, and executing the currently most viable action determined by the planning loop.
\end{itemize}
Online planning with its two loops is informally shown in Figure \ref{fig:olp}.

\subsection{Open Loop Planning}
Open loop planning is an approach to determine policies optimized w.r.t. some objective without storing information about the states that are intermediately encountered while simulating policy execution for evaluation purposes \cite{bubeck2010open,weinstein2013open}. Given a set of actions $A$, we are only interested in finding a plan $p \in A^h$, and we are only keeping information about the action sequences in order to guide the planning process. That is, we reformulate the policy to $\Pi : P(A^h | S)$, now being a distribution of sequences of length $h$ given a current state $s \in S$.

Open loop planning contrasts with closed loop planning, such as e.g. Monte Carlo Tree Search \cite{chaslot2010monte}, where action selection is typically conditioned by the history of previously encountered states and executed actions.

\subsection{Multiarmed Bandits}
Multiarmed bandits (MAB) are a core framework for decision making. A bandit consists of a number of arms, each representing an agent's choice. In our setting, each arm represents an action $a \in A$. Each arm provides a particular payoff, and the agent's goal is to identify the most preferable arm. It can explore the bandit by pulling one arm at a time, and observe the corresponding payoff.

An MAB can be interpreted as a simple Markov decision process with a single state. In their basic formulation, MABs already provide a clear framework for studying the exploration-exploitation tradeoff inherent to decision making under uncertainty: Should the agent select the arm that previously showed to be most promising? Or should it go on exploring other options? 
For a recent survey of MAB and its variants, see \cite{kuleshov2014algorithms}.

\subsection{Thompson Sampling}
\label{sec:ts}

Thompson sampling (TS) is a Bayesian algorithm for solving an MAB. It was proposed decades ago \cite{thompson1933likelihood}, but only recently its astonishing effectiveness and generality have been identified \cite{ortega2009bayesian,chapelle2011empirical,kaufmann2012thompson}.

TS infers a posterior distribution over $p$ based on the observed arm payoffs and a prior assumption about the distribution of $p$. In general, the posterior is proportional to the likelihood of observed data $D$ (i.e. an arm's observed payoffs), multiplied by the prior distribution $P(\theta)$ over the parameters of interest, $\theta = p$ in our case (Equation \ref{eq:bayes}).

\begin{equation}
\label{eq:bayes}
P(\theta | D) \propto P(D | \theta) P(\theta)
\end{equation}

%

TS maintains such a distribution $P(\theta)$ for each arm. The algorithm then samples a potential value for each arm from these distributions. It then plays the arm from whose distribution the maximum value has been sampled, observes the payoff, and uses this observation to update the corresponding distribution. Repeating this process results in almost sure identification of the arm with the highest payoff. TS is schematically shown in Algorithm \ref{alg:ts}.

\begin{algorithm}
	\begin{algorithmic}[1]
		\Procedure{Thompson Sampling}{}
		\State $\forall a \in A : \hat{v}_a \sim P_a(\theta)$
		\State play $\arg \max_a \hat{v}_a$ and observe result
		\State update $P_a(\theta)$ w.r.t. result
		\EndProcedure
	\end{algorithmic}
	\caption{Thompson Sampling}
	\label{alg:ts}
\end{algorithm}



\section{Distributed Online Open Loop Planning}
\label{sec:doolp}

In this Section, we describe the general DOOLP framework for multiagent coordination and decision making under uncertainty.

\subsection{Overview}

Distributed Online Open Loop Planning (DOOLP) realizes multiagent coordination by distributed simulation-based open loop planning. Each agent maintains a sampling policy that balances the exploitation-exploration tradeoff it faces when searching for an individual high-quality solution. Individual search is also influenced by other agents' current sampling strategies: Before simulating the consequences of a plan choice, an agent queries other agents it wants to coordinate with for a sample from their current policies. These queried plan samples are distributed w.r.t. the other agents' current preferences. As plan sampling, simulation and updating of sampling policies are performed iteratively, the other agents' samples influence the change of the individual policy by the degree of coordination that is present in expectation over the joint policy.

\begin{figure}
	\centering
	\includegraphics[width=\textwidth]{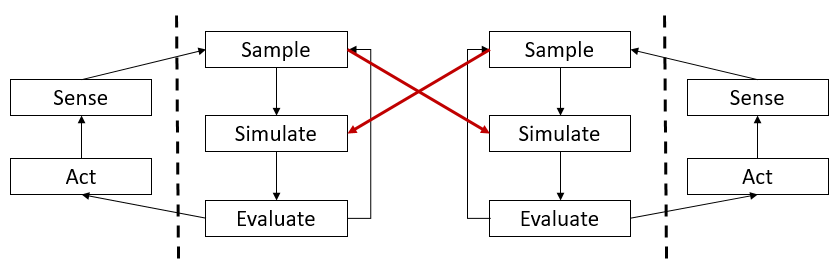}
	\caption{Schematic representation of distributed online open loop planning. The dashed vertical lines separate physical and cognitive components of an online planning agent. Two coordinating agents are shown. Their respective coordination by means of communication is highlighted in red. Communication is part of the high-frequency planning loop of each agent.}
	\label{fig:doolp}
\end{figure}

DOOLP is informally represented in Figure \ref{fig:doolp}: Two coordinating online planning agents are shown. Their respective coordination by means of communication is highlighted in red. We emphasize that communication is part of the high-frequency planning loop of each agent.

\subsection{DOOLP Formalization}

We formalize the DOOLP framework as follows. Note that, while the formalization is closely related to DecMDPs (cf. Section \ref{sec:related}), it is not necessary for DOOLP to specify a transition distribution and a reward function explicitly. Let the set of agents $Ag$, the state space $S$, the set of agent's actions $\{A_\textit{ag}\}$, the set of joint actions $A = \otimes_{\textit{ag} \in \textit{Ag}} A_\textit{ag}$ and the planning horizon $h$ are defined as for a DecMPD (cf. Section \ref{sec:related}). A DOOLP agent is a tuple $\mathcal{D} = (\Pi_\textit{ag}, C, \Delta)$, with the following definitions.
\begin{itemize}
	\item Each agent maintains a policy $\Pi_\textit{ag} : P(A_\textit{ag}^h | S)$. The policy is a probability distribution over plans (i.e. sequences of individual actions) given a state. This distribution is used to sample plans for which to simulate the consequences. The policy is updated after simulation in order to increase the probability of generating high-value plans.
	\item Each agent maintains a set of coordinating agents (e.g. neighbors, trusted agents, etc.) via a communication strategy $C : S \rightarrow 2^\textit{Ag}$. At each sampling step, the joint plan to be simulated is constructed from (a) sampling from the own current policies and (b) querying all currently coordinating agents in $C(s)$ for a plan sample from their respective current individual policies.
	\item Each agent maintains a simulation of domain dynamics $\Delta : P(S \times \mathbb{R} | S \times A)$. This simulation is a probability distribution capturing aleatoric uncertainty of domain dynamics, e.g. stochastic action effects. Given a current state and a joint action it outputs a successor state and a reward associated with the generated transition. The reward is a signal about the quality of the observed transition w.r.t. an agent's goals. DOOLP aims at maximizing the expected cumulative reward when executing a policy. Note that $\Delta$ mitigates the need to explicitly specify a DecMDP for a DOOLP agent. That is, the transition distribution and the reward function do not have to be explicitly formulated, but rather it is sufficient to provide a simulation of the application domain in order to apply DOOLP for distribution online coordination.
\end{itemize}

\subsection{DOOLP Planning and Execution}

Based on the DOOLP specification discussed above, we now describe the planning and execution loops for each agent.

\subsubsection{Planning}

DOOLP's inner high-frequency planning loop operates by repeatedly performing the following steps (cf. Algorithm \ref{alg:doolp_plan}).
\begin{itemize}
	\item A planning agent samples an individual plan $p \in A^h$ from its own individual policy $\Pi_\mathrm{self}$ (line 2).
	\item The agent then queries all other agents in its set of coordinating agents $C(s)$ for samples from their current policies $\Pi_c, c\in C(s)$ (line 3). We emphasize that this communication in combination with a sensible exploration-exploitation strategy implicitly yields coordination of multiple agents. Note that in general $C(s)$ is unrestricted and may indeed contain all other present agents. For efficiency purposes, $C(s)$ may also be constructed by other means, e.g. based on topological neighborhood, trust values, etc. 
	\item The agent simulates the joint plan built from its own plan and the coordinating agents' plans with the available mode of domain dynamics $\Delta$ and observes the associated rewards (lines 4 -- 7).
	\item The agent computes the cumulative reward for each step in the simulation (i.e. the cumulative sum of future rewards from any given planning step) and updates its individual policy based on the cumulative rewards in a way that increases the probability of generating high-value plans (i.e. plans whose expected reward is high) (lines 8 -- 11).
\end{itemize}

\begin{algorithm}
	\begin{algorithmic}[1]
		\Procedure {plan}{$s$}
		
%

		\State $p \sim \Pi_\mathrm{self}(s)$
		\Comment{sample plan from policy}
		
		\State $p_\mathrm{joint} \gets p \cup \left( \bigcup_{c \in C(s)} p_c \right)$
		\Comment{query other agents' plans and join}
		
		\State $r \gets \mathrm{nil}$
		\Comment{initialize plan rewards}
		
		\For {$a_i \in p_\mathrm{joint}$}
			\State $s, r_i \sim \Delta(s, a_i)$
			\Comment{simulate joint actions}
			\State $r \gets r :: r_i$
			\Comment{store reward}
		\EndFor
		
		\For {$h - 1 \geq i \geq 0$}
			\State $r_i \gets r_i + r_{i + 1}$
			\Comment {cumulative reward}
		\EndFor
		
		\For {$a_i, r_i \in p, r$}
			\State update $\Pi_\mathrm{self}$ w.r.t. $(a_i, r_i)$
			\Comment {update individual policy}
		\EndFor
		
		\EndProcedure
	\end{algorithmic}
	\caption{DOOLP planning loop}
	\label{alg:doolp_plan}
\end{algorithm}

\subsubsection{Execution}

DOOLP's outer execution loop repeatedly performs the following operations (cf. Algorithm \ref{alg:doolp}):
\begin{itemize}
	\item The current state is observed (line 3).
	\item The planning loop is executed until a user-defined event interrupts the loop, e.g. when a certain simulation budget has been reached, or an external event requires an agent's action (lines 4 and 5).
	\item The agent then uses the mode of its current policy $\Pi_\mathrm{self}$ to construct the plan with the highest expected future reward, and executes its first action (lines 6 and 7).
\end{itemize}

\begin{algorithm}
	\begin{algorithmic}[1]
		\Procedure{DOOLP}{}
		
		\Loop
		
		\State observe current state $s$
		
		\While{not interrupted}
		
		\State \Call{plan}{$s$}
		
		\EndWhile
		
		\State $p \gets \mathrm{mode}(\Pi_\mathrm{self})$
		\State execute $p_0$
		
		\EndLoop
		
		\EndProcedure
	\end{algorithmic}
	\caption{DOOLP execution loop}
	\label{alg:doolp}
\end{algorithm}

We assume communication of a DOOLP agent (listening and answering queries) to run in parallel to the execution loop. We assume this communication routine to be non-blocking, and to be able to access the current state and the policy of the agent in order to sample and return a currently viable plan to any querying agent.

\section{Distributed Online Thompson Sampling}
\label{sec:dots}

DOOLP is instantiated by implementing the following operations:

\begin{enumerate}
	\item Representation of the policy $\Pi_\mathrm{self}$.
	\item Sampling plans from the policy.
	\item Updating the policy given observed action-reward tuples.
\end{enumerate}

In the following, we describe each of these points for distributed online Thompson sampling as a DOOLP instance.

\subsection{Policy Representation}
\label{sec:policy}

DOTS uses a stack of multiarmed bandits to represent the policy $\Pi_\textit{ag} : P(A_\textit{ag}^h | S)$ of an agent $\textit{ag} \in \textit{Ag}$. For each planned step, DOTS maintains a bandit that models an agent's belief about action values for the corresponding step in the plan generated by the policy, based on previous simulations of plans via $\Delta$.

\subsubsection{Action Values}
By simulating execution of an action sequence from an initial state with $\Delta$, an agent obtains a sequence of states, actions and rewards $s_1, a_1, r_1, ..., s_h, a_h, r_h$. Given such data, we define the value of an action $a_i, 1 \leq i \leq h$ as the cumulative reward gained onwards from executing that action.
$$\hat{V}(a_i) = \sum_{i \leq j \leq h} r_j$$

In order to estimate an action's expected value, DOTS maintains a buffer of observed action sample values $X_{a,i}$ for each action $a \in A$ at each step $i \leq h$. As multiagent coordination yields a moving target distribution of values (i.e. concept drift occurs in the process of coordination), we maintain the buffer in a sliding window fashion.

\subsubsection{Bayesian Estimation of Action Value Expectation}
Given a buffer of observed action values $X_{a,i}$ for an action $a$ at depth $i$, an agent can estimate the corresponding action's value distribution in a Bayesian way as described in the following.
DOTS assumes that the expectation of an action's value is normally distributed with mean $\mu$ and precision $\tau$.
\begin{equation}
x \sim \mathcal{N}(\mu, \tau^{-1})
\end{equation}

These parameters are unknown initially, and are to be estimated by DOTS based on simulation of action sequences with $\Delta$.
We place a normal-gamma prior over the parameters $\mu$ and $\tau$ of this distribution to model an agent's initial uncertainty about $\mu$ and $\tau$. The normal-gamma prior is parametrized by a prior mean $\mu_0$, the number of prior mean pseudo-observations $\lambda_0$, the number of prior variance pseudo-observations $\alpha_0$ and $\beta_0$, such that $\frac{\beta_0}{\alpha_0}$ is the prior's sample variance. Given these parameters, we can sample prior mean and precision from the corresponding normal-gamma distribution.
\begin{equation}
(\mu, \tau) \sim \mathcal{NG}(\mu_0, \lambda_0, \alpha_0, \beta_0)
\end{equation}

The normal-gamma posterior parameters of the action value distribution $\mu$ and $\tau$ after observing action values $X = {x_1, ..., x_n}$ are computed as follows.
\begin{align}
P(\mu, \tau | X) &= \nonumber \\
\mathcal{NG} &\left(\frac{\lambda_0 \mu_0 + n \bar{x}}{\lambda_0 + n}, \lambda_0 + n, \alpha_0 + \frac{n}{2}, \beta_0 + \frac{1}{2} \left( ns + \frac{\lambda_0 n (\bar{x} - \mu_0)^2}{\lambda_0 + n}\right)\right)
\label{eq:ng}
\end{align}

Here, $n = |X|$ is the number of observations, $\bar{x} = \frac{1}{n} \sum_{x_i \in X} x_i$ is the observations' mean, and $s = \frac{1}{n} \sum_{x_i \in X} (x_i - \bar{x})^2$ is the observations' sample variance.

\subsection{Sampling from the Policy}
Sampling a plan from the policy $\Pi_\mathrm{self}$ (cf. Algorithm \ref{alg:doolp_plan}, line 2) is performed via Thompson sampling (cf. Section \ref{sec:ts}). Let $h \in \mathbb{N}^+$ be the planning depth. For each planning step $i \in \{0, ..., h\}$ and each action $a \in A$, the agent has observed the rewards-to-go $X_{a,i}$. To sample a plan from $\Pi_\mathrm{self}$ an agent builds a normal-gamma distribution $P(\mu_{a,i}, \tau_{a,i})$ from the corresponding previously observed $X_{a,i}$ as defined by Equation \ref{eq:ng}. It then samples a parametrization of the reward distribution for each planning step $i \in \{0, ..., h\}$ and each action $a \in A$. The sequence of actions maximizing $\mu_{a,i}$ for each $i$ then form the sampled plan. Algorithm \ref{alg:sampling} shows the corresponding sampling procedure.

\begin{algorithm}
	\begin{algorithmic}[1]
		\State $p \gets \mathrm{nil}$
		\For{$0 \leq i < h$}
		\State $\forall a \in A_\mathrm{self} : (\mu_{a,i}, \tau_{a,i}) \sim P( \cdot | X_{a,i})$
		\State $a_i \gets \arg \max_a \mu_{a,i}$
		\State $p \gets p :: a_i$
		\EndFor
	\end{algorithmic}
	\caption{Sampling a plan from the policy $\Pi_\mathrm{self} = P( \cdot | X_{a,i})$ with Thompson sampling, given observed rewards-to-go $X_{a,i}$. For the DOTS algorithm, this procedure implements line 2 in Algorithm \ref{alg:doolp_plan} of the DOOLP framework. It is also used to sample plans for answering communication queries from other agents.}
	\label{alg:sampling}
\end{algorithm}

\subsection{Updating the Policy}
Updating the policy is implicitly done by updating the set of observed action values $X_{a,i}$ for actions that are simulated in the planning loop (cf. Algorithm \ref{alg:doolp_plan}). Changing $X_{a,i}$ directly influences the distributions $P(\mu_{a,i}, \tau_{a,i} | X_{a,i})$, which in turn define the density of sampled plans (cf. Algorithm \ref{alg:sampling}, line 3).

As multiagent planning yields a moving target (due to agents concurrently changing/adapting their preferences), we treat the $X_{a,i}$ in a sliding window fashion in order to only reflect the most recent observed/simulated action values for a particular action choice. That is, all $X_{a,i}$ are implemented as buffers in a first-in, first-out manner, only keeping track of the most recent evaluations.

\section{Case Study}
\label{sec:results}

We empirically evaluated the effectiveness of DOTS in a smart factory case study.

\subsection{Domain Setup}

We considered a setting where various items are to be processed in a smart factory consisting of a number of different machines. Each item carries constraints on the type of processing that it has to pass in order to proceed, and also on the order of the processing steps. Each machine is associated with a processing type, a processing cost and a processing failure probability. The latter is exemplary for domain inherent stochasticity.

When not enqueued at a machine, agents decide at which machine to enqueue in order to get their processing steps done as fast as possible. This implicitly requires coordination of requests and resources, as a machine can only process one item at a time. Agents also have the option to do nothing, i.e. to wait.
The reward generated by the domain depends on the number of processed request, as well as the processing cost associated with the processing machines.

Note that the resulting problem grows exponentially with the number of agents. For a setting with $i$ items and $m$ machines, and plans of length $|p|$, the resulting search space is of cardinality $(m + 1)^{i^{|p|}}$. This means that in a setting with 8 items, 4 machines and a plan length of four there are more than $10^{22}$ joint plan options for the items in any given situation. Also, as machines have a Bernoulli failure probability, there is a very high branching factor regarding the consequences of joint actions.

To ease reproducibility, an implementation of our experimental setup can be downloaded from \url{https://github.com/jazzbob/doolp}.

\subsection{Coordination Variants}

We evaluated DOTS in our setup by comparing it to three baselines.

\begin{itemize}
	\item As a first baseline, we used a distributed random search approach, that we label Vanilla Monte Carlo (VMC). For VMC, each agent samples potential plans uniformly at random. There is no sampling policy update step of the policy due to observed simulation rewards. The joint sampled plans are then simulated and each agent keeps track of the plan that achieved the best joint value so far. The first action of the best found plan is executed when the planning loop is interrupted, and the process repeats.
	Note that VMC does not update its sampling policy based on observed action-reward tuples. Applying VMC in a multiagent scenario can therefore be interpreted as implicit coordination. I.e., with VMC coordination only occurs due to the resulting states the system encounters at runtime, but not due to a dedicated coordination effort.
	
	\item As a second baseline, we used an instantiation of the DOOLP framework where sampling is performed in an $\epsilon$-greedy manner. I.e., with probability $1 - \epsilon$, the action with maximum mean previously observed rewards-to-go is selected. With probability $\epsilon$, a random action is sampled uniformly from the action set for exploration purposes. We set $\epsilon$ to 0.1 in our experiments. This DOOLP instantiation updates the sampling policy by building the mean observed cumulative reward for each action at each planning step. In contrast to the VMC baseline, this baseline performs explicit coordination as specified by DOOLP. However, in contrast to DOTS, it does not model uncertainty about its cumulative reward estimates, but rather uses a maximum likelihood approach for value estimation.
	
	\item As a third baseline, we used an instantiation of the DOOLP framework where sampling is performed with the UCB algorithm \cite{auer2002finite}. UCB is a well-known selection strategy for multiarmed bandits based on upper confidence bounds of the reward expectation for each arm, yielding an exploration behavior known as \textit{optimism in the face of uncertainty} \cite{bubeck2010open}. Let $n$ be the number of a multiarmed bandit has been sampled, and let $n_a$ be the number of samples for a particular action $a$. The UCB score for an action $a$ is given by the following term.
	$$\mathrm{UCB}_a = \bar{x}_a + c \cdot \sqrt{\dfrac{2 \ln n}{n_a}}$$
	Here, $\bar{x}_a$ is the sample mean of observed rewards for action $a$ and $c > 0$ is a constant weighting the exploration term. We set $c = 1$ in our experiments. UCB selects the action that maximized the UCB score, i.e. $\max_{a \in A} \mathrm{UCB}_a$. Policy updating is done by keeping track of observed rewards and the number of samples for each action, directly influencing the UCB scores.
\end{itemize}

We evaluated DOTS in various settings, with similar results. Here we report on a setting with 8 items, 4 machines and a planning depth of 4. For each executed action a number $n \in \{64, 128, 256, 512\}$ of simulations were performed (i.e. planning loop interruption occurred after $n$ simulations). Note that in all cases, $n \ll 10^{22}$, that is, only a fraction of the joint plan space is searched. We used a sliding window size of 10 for the size of the rewards-to-go buffers $X_{a,i}$.

We used the following normal-gamma prior to model initial agent uncertainty about action value distributions (cf. Section \ref{sec:policy}).
\begin{align}
	\centering \nonumber
	\mu_0 = 0, \lambda_0 = 1, \alpha_0 = 1, \beta_0 = 100
\end{align}

Note that the influence of prior parameters on the posterior distribution is reduced with increasing numbers of observations used for inferring the posterior.

\subsection{Results}

We posed the following research questions to be answered by our experiments.
\begin{enumerate}
	\item Is DOOLP realizing coordination effectively by interleaving planning, communication and execution?
	\item Does DOTS' Bayesian uncertainty treatment yield a positive effect on coordination quality?
	\item Is DOOLP robust w.r.t. communication coverage, i.e. is it scalable to many communicating agents?
\end{enumerate}

\subsubsection{Coordination Effectiveness}

\begin{figure}
	\subfloat[64 planning iterations\label{img:reward64}]{%
		\includegraphics[width=0.5\textwidth]{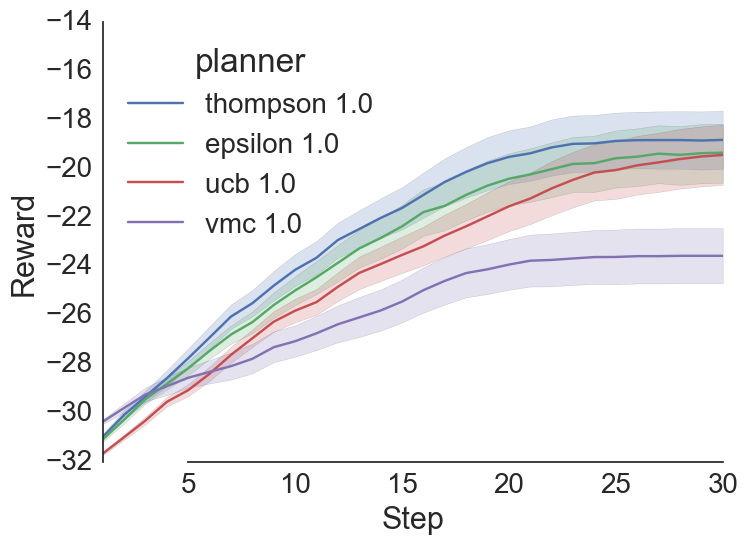}
	}
	\hfill
	\subfloat[128 planning iterations\label{img:reward128}]{%
		\includegraphics[width=0.5\textwidth]{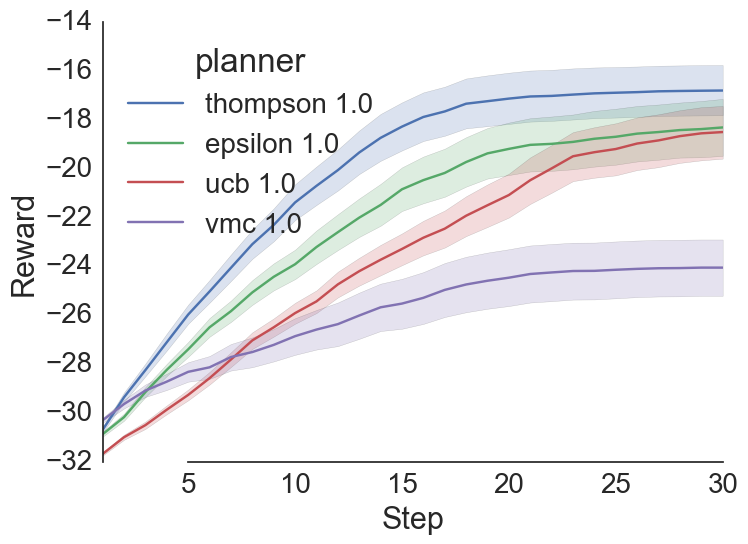}
	}
	\vfill
	\subfloat[256 planning iterations\label{img:reward256}]{%
		\includegraphics[width=0.5\textwidth]{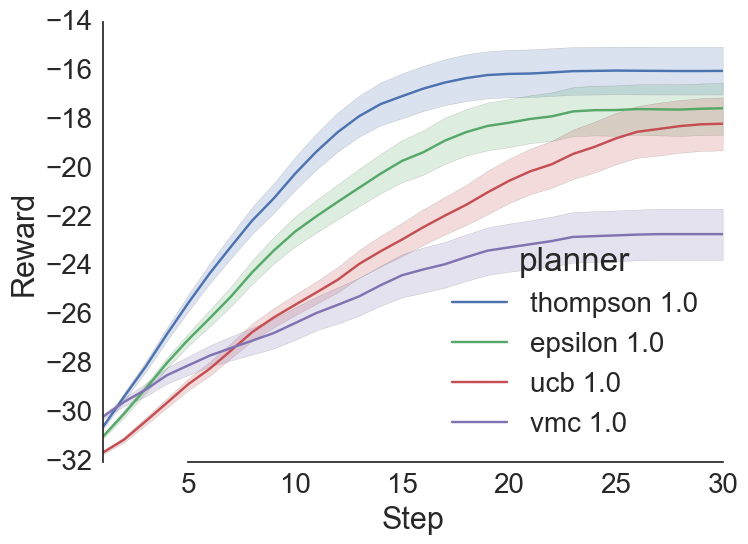}
	}
	\hfill
	\subfloat[512 planning iterations\label{img:reward512}]{%
		\includegraphics[width=0.5\textwidth]{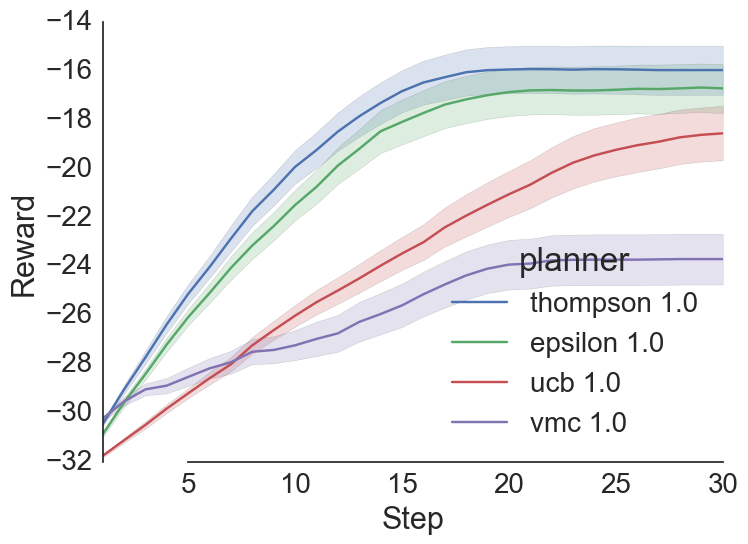}
	}
	\caption{Average achieved rewards (50 runs) over time for different planner variants and planning iterations per executed action. Shaded areas show one standard deviation. Best viewed in color.}
	\label{img:scores}
\end{figure}

Figure \ref{img:scores} shows the average cumulative scores w.r.t. discrete time steps (i.e. number of execution loops) achieved by the agents planning their actions in a distributed manner with DOTS. Also shown are results of the three baseline approaches (VMC, $\epsilon$-greedy, UCB). The shaded areas show one standard deviation of the results. DOTS consequently outperformed the baseline approaches regardless of the number of simulations and policy updates performed before executing agents' actions. The $\epsilon$-greedy variant of DOOLP was able to reach the performance of DOTS for 512 planning iterations. UCB performed weaker than both in all settings. As VMC yields significantly lower rewards in all settings than all DOOLP variants, we conclude that explicit coordination is indeed realized by instantiations of DOOLP. Given our observations, we give a positive answer to questions 1 and 2 above.

\subsubsection{Coordination Robustness and Scalability}
We evaluated the robustness and scalability of DOTS w.r.t. to communication coverage. These are important characteristics of multiagent planning, as communication uses possibly scarce or expensive resources (e.g. bandwidth), and is typically prone to failure (e.g. message loss). For each iteration of the planning loop, we randomly dropped a fraction of agents from the communication set $C$, resulting in gossip communication between coordinating agents \cite{shah2009gossip}. This also reduces the computational resources needed for simulation, as only the remaining agents in $C$ were participating in queried joint plans.

\begin{figure}
	\subfloat[DOTS\label{img:reward_thompson}]{%
		\includegraphics[width=0.5\textwidth]{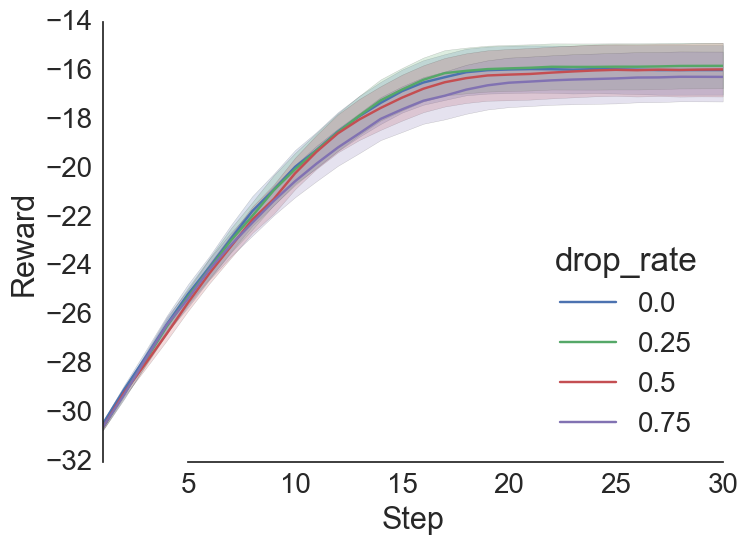}
	}
	\hfill
	\subfloat[Epsilon\label{img:reward_epsilon}]{%
		\includegraphics[width=0.5\textwidth]{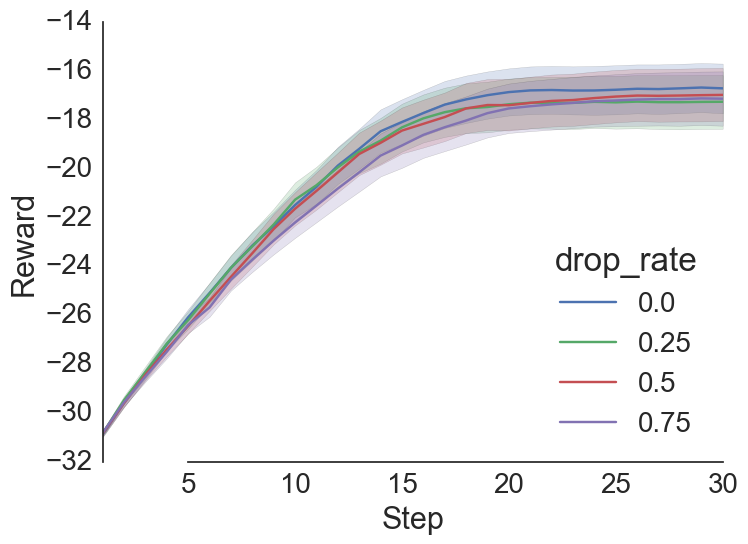}
	}
	\vfill
	\subfloat[UCB\label{img:reward_epsilon}]{%
		\includegraphics[width=0.5\textwidth]{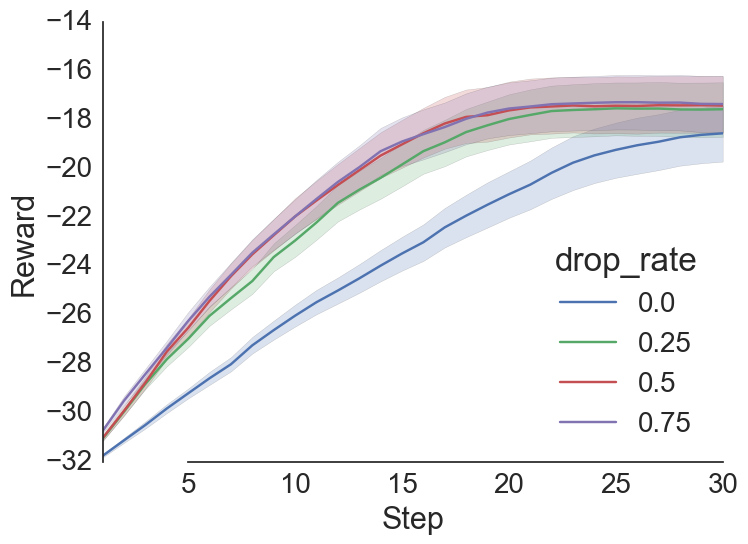}
	}
	\hfill
	\subfloat[VMC\label{img:reward_vmc}]{%
		\includegraphics[width=0.5\textwidth]{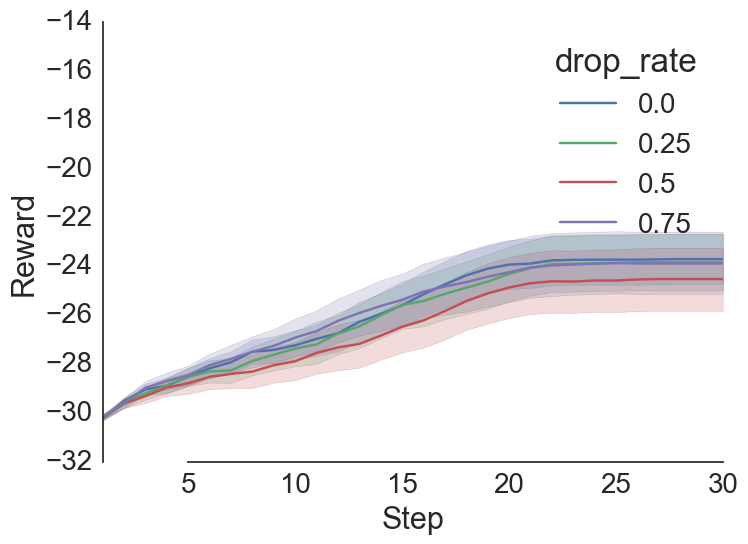}
	}
	\caption{Average achieved rewards (50 runs) over time for different planner variants and communication drop rates. Shaded areas show one standard deviation. 512 planning iterations per executed action. Best viewed in color.}
	\label{img:gossip}
\end{figure}

Figure \ref{img:gossip} shows the corresponding comparison of average achieved cumulative scores for different communication drop rates (i.e. fractions of agents not included in the joint plans that are simulated) for 512 planning iterations. We measured drop rates of $0, 0.25, 0.5$ and $0.75$. We observe that DOTS and $\epsilon$-greedy DOOLP are robust against limiting communication between agents. Interestingly, the UCB variant increased its performance when reducing communication. We conjecture that this is due to the optimistic nature of UCB sampling. It is known that early optimism may result in poor local optima in multiagent coordination \cite{panait2008theoretical}. UCB sampling may emphasize this effect, and reducing communication possibly mitigates it. As expected, VMC does not suffer significantly from reduced communication as agents guess their individual plans regardless of communication (i.e. there is no sampling strategy update due to communication). We observed similar results for fewer planning iterations, with DOTS becoming slightly sensitive to communication drop rates, still yielding the best results of all compared DOOLP variants in all cases (see Appendix \ref{sec:appendix_a} for results). Given our results, we give a positive answer to question 3 above.

\paragraph{}
We conclude that explicit coordination with the DOOLP framework provides a scalable and efficient way for multiagent coordination under uncertainty, and that DOTS' Bayesian modeling of action value uncertainty yields additional coordination performance in comparison to maximum likelihood estimation as done by the $\epsilon$-greedy baseline and action selection based confidence bounds as done by the UCB variant.

\section{Conclusion \& Further Work}
\label{sec:conclusion}

We proposed \textit{distributed online open loop planning} (DOOLP), a framework for scalable online multiagent coordination and decision making under uncertainty. We also proposed a particular instantiation of DOOLP, \textit{distributed online Thompson sampling} (DOTS). DOTS uses a Bayesian approach for modeling uncertainty in order to achieve coordination in cooperative multiagent settings. We have presented a formal description of DOOLP and DOTS, and evaluated its effectiveness empirically on a smart factory case study. We also evaluated the robustness of various DOOLP variants w.r.t. communication rates between agents, and observed a highly robust coordination quality w.r.t. communication rates between coordinating agents. Our results show that DOTS is a viable candidate for robust and scalable online multiagent coordination under uncertainty.

The DOOLP framework could straightforwardly be extended to more complex settings, including asynchronous coordination, local agent knowledge, heterogeneous reward functions, actions duration planning \cite{belzner2016time} or different optimization objectives such as risk metrics or quality of service (see e.g. \cite{rockafellar2000optimization,belzner2016qos}).

Another direction would be to incorporate simulation models learned from data (e.g. runtime observations), and to deal with model uncertainty arising from the learning process in the planning process. Also, statistical system verification under these constraints is a direction of current research. See e.g. \cite{belzner2017a} for recent work of the authors in this direction. 

Another interesting venue for future research is to integrate global, emergent phenomena into the local planning and coordination efforts. This could for example be achieved by learning predictive models (e.g. \cite{schmidhuber2015deep,bengio2015deep}) about global effects of interest (e.g. safety or quality w.r.t. global system requirements), emerging from individual interaction of all agents. The learned predictive model could in turn be used to guide the local planning processes \cite{hester2013texplore,silver2016mastering}.

\bibliographystyle{splncs}
\bibliography{references}

\appendix
\section{Coordination Robustness and Scalability for Different Numbers of Planning Iterations}
\label{sec:appendix_a}

\begin{figure}
	\subfloat[DOTS]{%
		\includegraphics[width=0.24\textwidth]{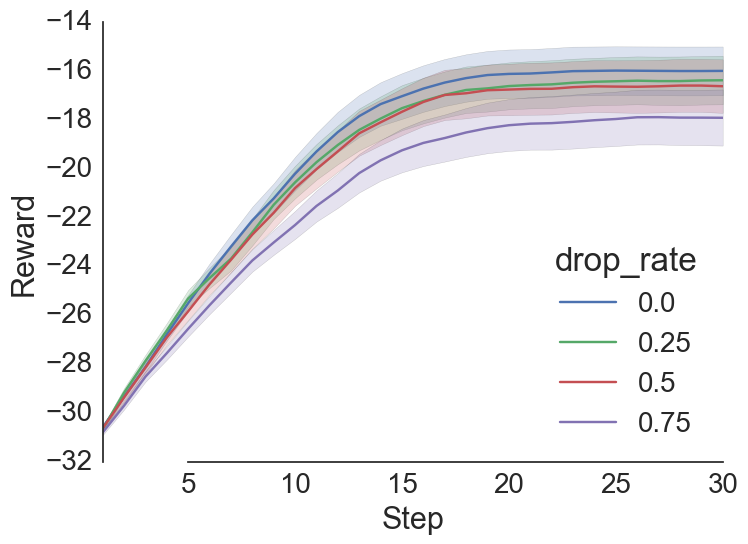}
	}
	\hfill
	\subfloat[Epsilon]{%
		\includegraphics[width=0.24\textwidth]{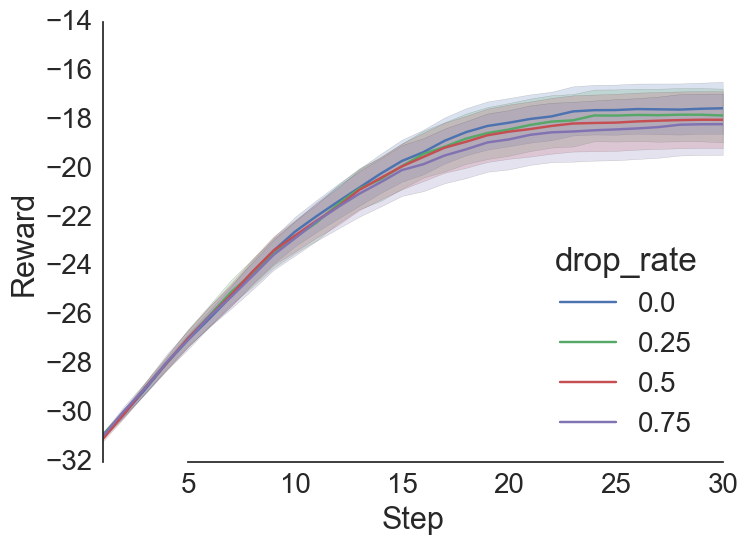}
	}
	\hfill
	\subfloat[UCB]{%
		\includegraphics[width=0.24\textwidth]{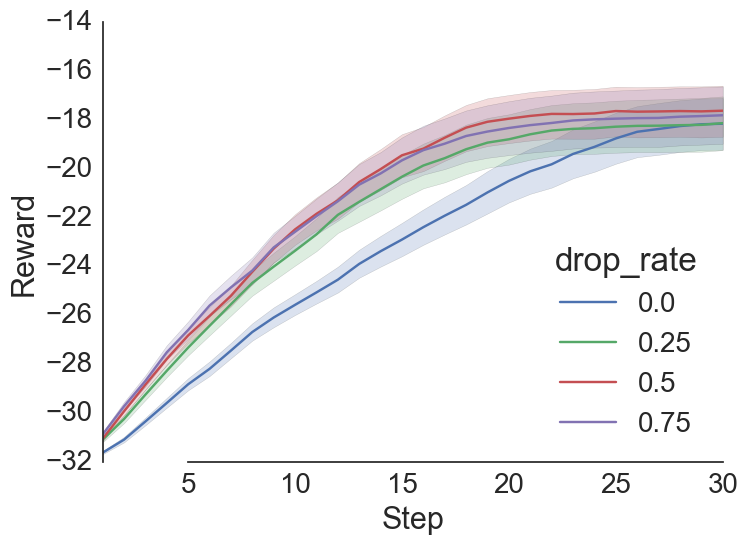}
	}
	\hfill
	\subfloat[VMC]{%
		\includegraphics[width=0.24\textwidth]{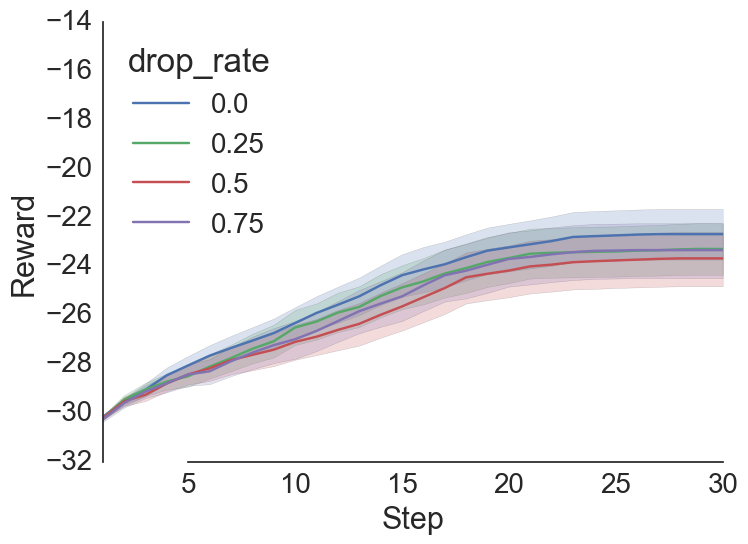}
	}
	\caption{Average achieved rewards (50 runs) over time for different planner variants and communication drop rates. Shaded areas show one standard deviation. 256 planning iterations per executed action. Best viewed in color.}
	\label{img:gossip256}
\end{figure}

\begin{figure}
	\subfloat[DOTS]{%
		\includegraphics[width=0.24\textwidth]{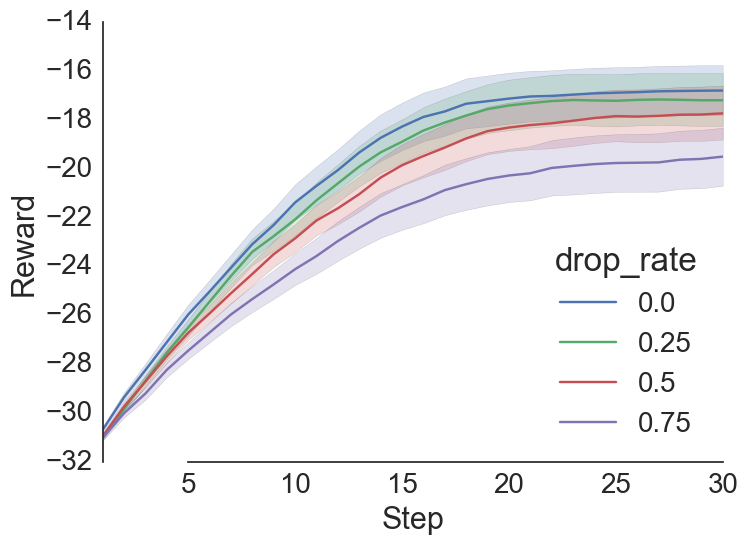}
	}
	\hfill
	\subfloat[Epsilon]{%
		\includegraphics[width=0.24\textwidth]{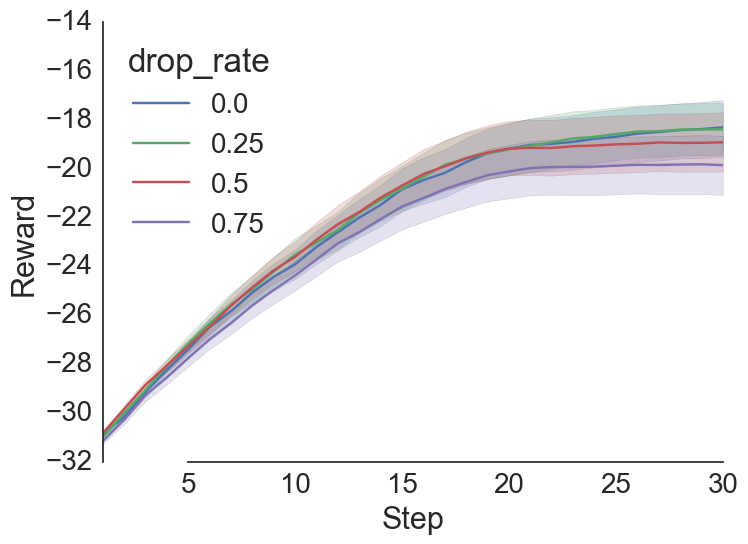}
	}
	\hfill
	\subfloat[UCB]{%
		\includegraphics[width=0.24\textwidth]{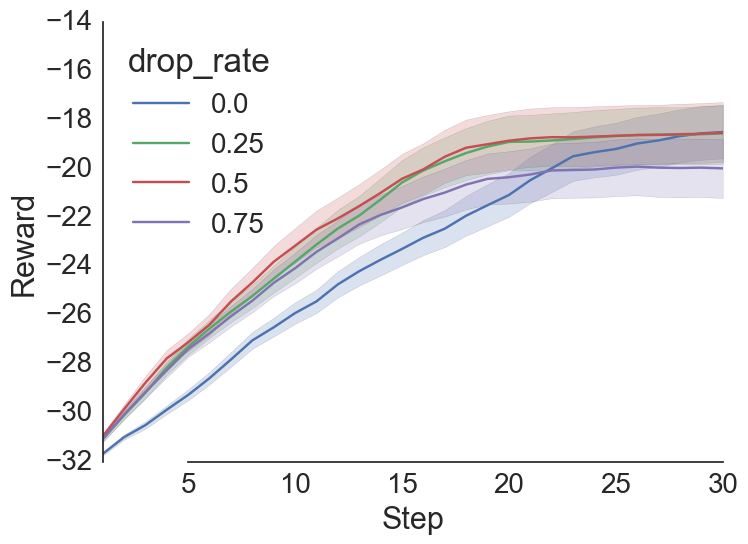}
	}
	\hfill
	\subfloat[VMC]{%
		\includegraphics[width=0.24\textwidth]{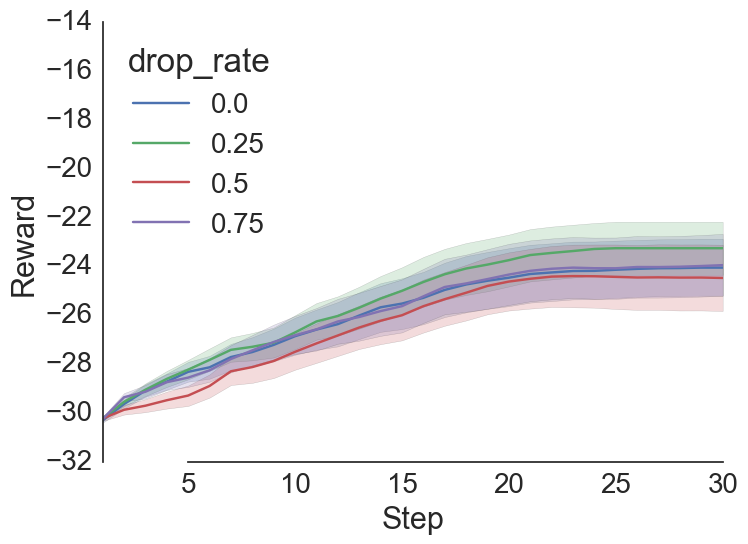}
	}
	\caption{Average achieved rewards (50 runs) over time for different planner variants and communication drop rates. Shaded areas show one standard deviation. 128 planning iterations per executed action. Best viewed in color.}
	\label{img:gossip128}
\end{figure}

\end{document}